\documentclass[11pt]{article}
\usepackage{graphicx,amsmath,amsfonts,amssymb,dcolumn,amsthm,euscript}
\def\bea{\begin{eqnarray}}
\def\eea{\end{eqnarray}}
\def\be{\begin{equation}}
\def\ee{\end{equation}}\usepackage{color}

\newcommand\nn{\nonumber} 
\newcommand\eq{\end{equation}}
\newcommand\bq{\begin{equation}}

\def\bar{\overline}
\newcommand\pa{\partial}

\def\d{{\rm d}}

\def\demi{\frac{1}{2}}

\begin{document}
\title{\textbf{ On  Stochastic Quantisation of Supersymmetric Theories  
}}
\author{\textbf{Laurent~Baulieu$^\dagger$
}
\thanks{{\tt baulieu@lpthe.jussieu.fr}
}
\\\\
\textit{$^\dagger$ LPTHE, Sorbonne Universit\'e, CNRS} \\
\textit{{} 4 Place Jussieu, 75005 Paris, France}\\
}

\def\blue{  \color{blue}}
\def\red{  \color{red}}

\date{}
\maketitle
\begin{abstract}
We explain how stochastic TQFT supersymmetry can   be made compatible with space supersymmetry. Taking the case of   $N = 2$ supersymmetric quantum mechanics, (the proof  would be the same for the Wess--Zumino model), we determine  the kernels that ensure the convergence of the stochastic process toward the standard  path integral, under the condition that they  are covariant under supersymmetry.  They depend on a massive parameter $M$ that can be chosen at will and  modifies the course of the stochastic evolution, but the    infinite stochastic time limit of the correlation functions  is in fact independent on the  choice of $M$.
\end{abstract}

\def\s{{s_{stoc}}}
\def\hl{{\hat l}}

\def\l{\lambda}
\def\ll{{\hat \lambda}}
\def\th    {\theta}
\def\thh{{\hat \theta}}
\def\ee{{\epsilon}}
\def\eeh{{\hat \epsilon}}
\def\dd{\delta}
\def\ddh {\hat \delta}

\def \pa{\partial}
\def\l{{\lambda}}
\def\g{{\gamma}}
\def\T{{t}}
\def\t{\tau}
\def \d{\delta}
\def\d{{\rm d}}
\def\x{{q}}
\newpage

\section{introduction}{
Giving the existence  of stochastic quantisation, an  interesting   problem  is to understand the way    stochastic evolution can preserve space supersymmetry, or,  more mathematically,  how   physical  space supersymmetry and stochastic TQFT supersymmetry can be made compatible.

Stochastic quantisation  computes  the Euclidean quantum correlation functions of a   given field theory in a $d$-space $\{ x^\mu\} $  by interpreting   this space as the $\tau=\infty$  boundary of a 
$d+1$-space,  $\{ x^\mu     \} \to 
\{ x^\mu,\tau  \}$.    The stochastic time $\tau$ is in fact     a bulk coordinate and  stochastic quantisation gives a  microscopic  understanding of  the Euclidean path integral measure  $\exp -\frac{S}{\hbar}$ as an equilibrium distribution at $\tau=\infty$, analogously as Langevin equations determine  the thermal equilibrium Boltzmann formula~\cite{parisi}\cite{huffel}.
It is also  a fact that a BRST-TQFT localisation procedure can   enforce  Langevin equations in a supersymmetric way  so  that the whole apparatus of stochastic quantisation can be expressed as a TQFT in  $d+1$ dimensions~\cite{stoctqft}. This  gives a good handle on the arguments in \cite{gozzi}.   

The   drift force of  the   stochastic  process  is basically  proportionate to  the Euclidean equations of   motion of the theory on the boundary, modulo some other forces that can possibly improve the convergence  of the process.   It is often  the case that       kernels that realise   factors of multiplication of the equations of motion must be carefully chosen to improve, and, sometimes, to define,       the convergency of the stochastic process~\cite{parisi}.  For maintaining space supersymmetry, if any, appropriate kernels are needed. The goal of this note is to explain how they can be determined.

It is  natural  to ask  how  symmetries,  which are realised on the boundary at  $\tau=\infty$, are represented      in the bulk, that is, at finite values of the stochastic time.  For gravity and gauge symmetries, and even string theory, one solves the question rather elegantly    in the framework of equivariant cohomology ~\cite{stoctqft}\cite{bbs}.  It   leads one    to define   a stochastic equivariant  topological BRST invariance. One then obtains  a universal frame for  the  stochastic quantisation of any given  theory with a gauge invariance, including a method to fix the gauge invariance in the bulk with a BRST symmetry construction. 

The case of  theories with   space supersymmetry is puzzling. The  equations of motion have a different supersymmetry  covariance  than the fields. Thus,    they cannot  be  used as       drift forces without introducing appropriate kernels to modify  consistently   their covariance. In this note, we  study in details the case of theories with global supersymmetry.  We   take the case  of   the simple    $N=2$ supersymmetric mechanics  and explain  how  the  supersymmetry of the   theory on the boundary can be   enforced   in the bulk. The case of the Wess and Zumino model can be handled in exactly the same way, modulo some notational complications, and we let the reader   make the correspondence.  The  superfield formalism is handy to get  Langevin equations that are compatible with supersymmetry for all values of the stochastic time. The method  we follow to compute the appropriate kernels and enforce supersymmetry in the  stochastic bulk    can be   actually applied   to more delicate  cases,   including when supersymmetry mixes with  gauge symmetries, as will be shown in a separate publication.
\def\v{ {\varphi}}

\section{A reminder  of kernels in stochastic quantisation} 
Consider a quantum field theory  of   fields $\v_ a(x )$ with a local action $S[\v]= \int \d  x L(\v _a(x)) $, where $a$ is some index that labels the fields, and the $x$'s  denote   Euclidean coordinates in $d$ dimensions. 
Stochastic quantisation introduces the stochastic time  coordinate~$\tau$,  with $\v _a(x)\to  \v _a(x,\tau)$.  The    $\t$  evolution of  the field  $\v_ a(x,\tau )$  is defined by  a Langevin process,  such that the path integral weight $\exp -S[\v]$ is its equilibrium distribution, when it exists.  Given a correlation function   $<f(\v(x))>$  in the   space  $\{x^\mu\}=\{x^\mu, \tau=\infty\}$, which  can be computed by the usual Euclidean path integral on the   space $\{x^\mu\} $, the fundamental property  of stochastic quantisation is that  
\bea\label{es}
< f(\v)> \equiv \int [\d \v]_x  f(\v)  \exp-S[\v] =\lim_{\tau\to\infty}     <<f(\v ^\eta(x,\tau)>>^{K,S}
\eea
The $(x,\tau)$ dependent correlators    $<<f(\v ^\eta(x,\tau)>>^{K,S}$ 
are  computed as  an  average over the  noises $\eta_a    (x,\tau)$. Here,    $\v ^\eta $   is computed in function of  the noises $\eta _a$ by solving    the following Langevin differential equation with some any given initial  condition  $\v ^\eta(x,\tau=\tau_0)=
\v _0(x)$,
\bea \label{lang}
\frac {\d  \v  _a } {\d \tau} &=& -K_{ab} \frac {\delta    S     } {\delta  \v_b}  +\eta _a
\eea
The dependance in $\hbar$ is   through  a rescaling of the noise $\eta$.  The correlations of the noise are  in fact   defined by the following formula, true for any  given  sufficiently regular functional $f$   \bea\label{Langevin}
<<f (\v^\eta)>>^K\ &\equiv  &\int [\d \eta_a]_{x\tau} \ 
f( \v^\eta)  \exp -\frac{1}{2} \int \d x\d t \ \eta _a K_{ab}^{-1}  \eta_b
\eea 
 It  is often the case that $K_{ab} $ must be  different from  the trivial identity  to ensure  a well-behaved drift force in the r.h.s of the Langevin equation~(\ref{Langevin}), and   a proper behaviour of the stochastic time evolution
at $\tau\to\infty$. One must indeed have a uniformly attractive  force in the Langevin process for all components of fields. Thus $\eta$ is not necessarily a white noise, and 
this happens for instance for the quantisation of a   spinor \footnote{ For a free Dirac action  
$ 
S 
=   \int \d x      \bar  \l     (     \gamma  \cdot \pa -m   ) \l      
$  
one can use  the following  Langevin equation  
$
\dot \l
=    (\gamma  \cdot \pa  +m) (    (  \gamma  \cdot \pa  -m   ) \l     +\eta_\l = 
    (  \pa^2  -m   ) \l     +\eta_\l, 
$
with    fermionic noise   correlation  functions 
$
<<   \eta_\l (x,\tau)   \eta_\l (x,\tau) >>
= \delta     (x-x') \delta   (\t-\t' )   \frac{1}{ \gamma  \cdot \pa+m}
$. It
gives the correct result
$P^{K =\gamma  \cdot \pa  +m} (\l,\tau=\infty)
=\exp -\int \d x      \bar  \l     (     \gamma  \cdot \pa -m   ) \l      
$.
If,  instead, we    take $
\dot \l
=     (  \gamma  \cdot \pa  -m   ) \l     +\eta_\l 
$ the Langevin process is ill-defined. 
In the former case the eigenvalues of  the drift force  $ (  \pa^2  -m   ) \l$ are positive, and the Langevin process   converges. In the later  case,   the drift force is    $  (  \gamma  \cdot \pa  -m   ) \l     $,   with positive and negative eigenvalues, so that the limit $\tau=\infty$ cannot be reached.
}. 

The dependence  on  $K_{ab} $ evaporates  when one reaches the limit $\tau=\infty$.
Indeed, given 
  different kernels $K$  that ensure a  convergence of the stochastic process,       the value of     $ \lim_{\tau\to\infty}     <<f(\v  (x,\tau)>>^{K,S}$ is independent on the choice of $K$, and is  given by the standard part integral formula in the r.h.s.   of Eq.~(\ref{es}).

  This property  can be demonstrated    from  the Fokker--Planck equation that is  implied by    the $K$~dependent  Langevin equation.  Indeed,  the   Langevin equation~(\ref{lang}) implies the existence of    a (Fokker--Planck) kernel     $ P^{S,K}(\v  (x) ,\tau)$,  
  which  permits one to computes  the time evolution of equal time stochastic correlators as follows
\bea  <<\v ( x_1,\tau ), ..,  \v (x_n, \tau)>>   = \int\  [\d \v]_x   \  \v ( x_1),..,  \v ( x_n)
P^{S,K}(\v , \tau)
\eea
Here,  $P^{S,K}$ is the solution of the Fokker--Planck equation
\bea\label{FP}
\frac {\pa P^{S,K}(\v ,\tau)}{\pa \tau} 
  =
  \int \d  x
\frac 
{ \delta }{\delta \v _a x}
K_{ab}
\Big (
 \frac   {\delta }{  \delta  \v _b(x)}
+
\frac   {\delta   S }{  \delta \v _b(x)}
\Big )
P^{S,K}(\v  ,\tau)
\eea
This equation needs  an  initial condition, eg: 
$P^{S,K}(\v (x), \tau=\tau_0) =  \delta (\v  (x) -\v_0(x))$. 
For sufficiently regular theories, one  can check an exponential damping   $\sim O(\exp-\tau)$ in  the dependence on the initial condition $\v_0(x)$ when $\tau\to\infty$.  The details of the evolution depend on the choice of $K$, but not the limit.
If a stationary distribution   $\lim_{\tau\to \infty}  P^{S,K}(\v (x,\t)$  exists,  Eq.~(\ref{FP}) implies
\bea
P^{S,K}(\v ,   \tau=\infty)   = \exp -\int  S[\v]
\eea
independently of the choice of $K$.  
At finite $\tau$, the $K$-dependance of $P^{S,K}$   is, explicitly :
\bea
P^{S,K}(\v (x)  ,\tau )= \Big ( \exp\tau  \int \d  x
\frac 
{ \delta  }{\pa  \v_a(x)}
K_{ab}  (
 \frac   {\delta }{ \delta \v _b(x)}
+
\frac   {\delta   S }{ \delta \v _b(x)}
)  
\Big)
P^{S,K}(V, \tau=\tau_0  )
\eea
  This    standard result of statistical mechanics  can be simplified   by the use of stochastic TQFT supersymmetry~\cite{ps}{\cite{gozzi}. 
There is  indeed  a supersymmetric path integral that  expresses the stochastic process in the bulk  $
\{ x,\tau\}$, where  $\v_ a(x) \to \v _a(x,\tau )$~{\cite{parisi}\cite{gozzi}.  It  enlarges   the phase space of the fields as follows:
\bea
\v_a  (x)\to    \Big( \v_ a(x,\tau ),       \Psi_{  _a} (x,\tau ),   \bar  \Psi _{   a}(x,\tau ),      b_a(x,\tau  \Big )
\eea
The stochastic    supersymmetry acts on this enlarged space of fields and it  is defined by  a nilpotent  graded differential operator  $\s$~\cite{stoctqft}.
$\s$  is alike a    BRST topological symmetry operation,  which justifies the name stochastic BRST supersymmetry, where   $\v _a,\Psi_a,\bar \Psi_a, b_a $ are identified 
as  two trivial BRST multiplets under $\s$ and\bea \s \v={\Psi }, \quad \s  {\Psi }=0,\quad \s{{\bar \Psi} } =b, \quad \s b=0   
\eea
If there a gauge symmetry acting on   $\v_ a(x) $,  some refinements are necessary and one must  build an equivariant supersymmetry~\cite{bbs}. 
The physical conclusions remain the same   for the relaxation to a stable equilibrium and they justify    the gauge theory path integral formula on the boundary.

We now  indicate a   universal  formula that will be useful in the following.
It  leaves  aside the unessential indices $x^\mu$  that can be discretised. 
Given  a pair   of boson fields $ (m(\tau),n (\tau))$   and a pair of  fermion fields  $ ({{p(\tau)} }, {q(\tau)})$, or the reversed  case   where on interchanges bosons and fermions,  
  determinant formula   imply    that one has  indeed the following identity
\bea \label {identity}
\int  [\d m]_\tau   [\d n  ]_\tau  [ \d  {p}]_\tau[ \d q ]_\tau
\exp\int d\tau  \Big [ n  \ ( {\cal M}     (m)  + \alpha   ) \mp  {p} \frac {\delta  {\cal M}(m) } {\delta m} q\  \Big]
=1
\eea
where $ {\cal M} (m)$ is any  given local  functional of  $m$.  $\alpha$ is independent on $m,n,p,q$ and the formula holds true whatever $\alpha$ is. Defining   $\s m=q,\s q=0,\s p=n,\s n=0$  and $\s$, one has   
\bea \label{id}  \s \Big (    p(  {\cal M} (m) + \alpha ) \Big )
= n ( {\cal {M} }(m)  + \alpha  ) \mp   p  \frac {\delta  {\cal M} }{\delta m} {q} 
\eea  
 $\s $ acts as  a graded differential   operator, so $\s^2=0$ on any functional of $m,n,p,q$.

The identity~(\ref{identity}) is important, as it  explains the chain  of identities in the next equation~(\ref{stoctruc}).
Indeed,
replacing $\alpha\to \eta$,  $m\to \v,n\to b, p\to =\bar \Psi ,q\to  \Psi$ and $ {\cal M} (\v )\to    \frac {\d  \v    } {\d t}  +K  \frac {\delta  S } {\delta \v}$ in the functional identity~(\ref{identity}),  one can write the following succession of equations :
\bea\label{stoctruc}
&&  \quad\quad\quad\quad \quad\quad\quad\quad\quad\quad\quad\quad<<f(\v)>>^K 
=\int [\d \eta_a]_\tau \ 
f(\v) \  \exp -\demi \int \d t \ \eta _a K_{ab}^{-1}  \eta_b
\cr
&=&\int [d \eta_a][ \d \v _a] [  \d  {\Psi _a} ][\d{{\bar \Psi} _a}]
f(\v)      \delta (\frac {\d  \v_a   } {\d \tau} + K_{a b}     \frac {\delta  {  S} } {\delta \v_b}  -\eta_a)
\exp- \int   \d \tau \     [\demi\eta_a  K_{ab}^{-1}  \eta_b -{{\bar \Psi} _a}
( \frac {\d      } {\d \tau}\delta_{ac} +  \frac {\delta   } {\delta \v_c}    K_{ab} \frac {\delta  {  S} }{\delta \v_b})
{\Psi _c} )  ]
\cr
& =&\int  [ \d \v _a] [  \d  {\Psi _a} ][\d {{\bar \Psi} _a}]
f(\v)    
\exp -\int   \d \tau     [\demi (\frac {\d  \v_a   } {\d \tau} +K_{a c}\frac {\delta  {  S} } {\delta c})    K_{ab}^{-1}  (\frac {\d  \v_b   } {\d \tau} +K_{bd}\frac {\delta  {  S} } {\delta d}) 
-{{\bar \Psi} _a}
( \frac {\d      } {\d \tau}\delta_{ac} +  \frac {\delta   } {\delta \v_c}    K_{ab} \frac {\delta  {  S} }{\delta \v_b})
{\Psi _c} )  ]
\cr
&=&\int  [ \d \v_a ][ \d b_a ][  \d  {\Psi _a} ][\d {{\bar \Psi} _a}] 
f(\v)    
\exp - \int   \d \tau   [-\demi  b_aK_{ab} b_b
+ b     (\frac {\d  \v_a   } {\d \tau} +K_{a b}\frac {\delta  {  S} } {\delta  \v_b}) 
-{{\bar \Psi} _a}
( \frac {\d      } {\d \tau}\delta_{ac} + \frac {\delta   } {\delta \v_c}    K_{ab} \frac {\delta  {  S} }{\delta \v_b})
{\Psi _c} )   ]
\cr
&=&\int  [ \d \v_ a ][  \d  {\Psi _a} ][\d {{\bar \Psi} _a}] [\d  b_a]
f(\v)    
\exp - \int   \d \tau     \s     {{\bar \Psi} _a}   [\      -  \demi K_{ab}  b_a +
\frac {\d \v_ a   } {\d \tau} +K_{ab} \frac {\delta  {  S} } {\delta \v_b}    ]
\eea
These equalities hold modulo  normalising    factors for each line. 
The first line enforces   the definition of $<<f>>^K$ as an average over noises, using the Langevin equation
$\frac{ \d  \v_a   } {\d t} =- K_{a b}\frac {\delta  {  S} } {\delta  \v_b}+\eta_a$ and an arbitrarily given initial condition 
$\v(\tau_0)$ at some reference time $\tau_0$ that one must introduce  
to  solve ~$\v$ in function of~$\eta$. 
The last line teaches us that  the correlator  $<<f(\v)>>^K $ is    computed by the path integral of a  topological quantum field theory,  with a localising gauge function related to the Langevin equation. In other words,  the path integral of the    $\s$-exact Lagrangian   in the last  line defines the same  correlation functions as those computed either from the Langevin equations, or   from their corresponding     Fokker-Planck  equations (if one  restricts oneself to  equal stochastic time  correlators). This supposes that  the possible zero modes of the operators that stand between the ghosts $\Psi$ and $\bar\Psi$  are handled properly with periodic boundary conditions for   $\Psi$.

 The  Hamiltonian that   defines  the stochastic time  evolution is the  $\tau$ Legendre transform of   the    supersymmetric Lagrangian  expressed  in the third or  fourth    lines of Eq.~(\ref{stoctruc}), as generically  explained   in~\cite{gozzi}. 

Eq.~(\ref{stoctruc}) is valid  whether   $\v _a$ is       bosonic or   fermionic, so  it can applied  when  $\v _a$ stands for any one of the field 
components  $x,\l ,  \ll,   A$ of the superfield $X(t)$ that we will shortly introduce.


\def \pa{\partial}
\def\l{{\lambda}}
\def\g{{\gamma}}
\def\T{{t}}
\def\t{\tau}
\def \d{\delta}



In the stochastic  BRST-TQFT supersymmetric formulation, the proof of the K independence  of the $\t=\infty$ limit of correlators, if they exist, follows from the properties of determinants,   for   the  bosonic case   as well as for   the fermionic case     
for $\v$. 
\def\ee{\alpha}
\def\ee{ }
\def\eeh{\hat \alpha}
\def\eeh{}
\def\d {{\rm d}}
\def\a{\a}
\section{Supersymmetric kernels}
Let us now consider a supersymmetric theory with a set of auxiliary fields that gives  a closed system of transformations. We will  prove  that supersymmetry  is compatible with the    stochastic time evolution.

For the   notational simplicity of the proof  we consider the  $N=2$ supersymmetric quantum mechanics. One has   $\{x^\mu\} =\{t\}$,     two supersymmetry generators 
$\dd_i= (\dd,\hat \dd)$    and one multiplet 
(1,2,1), that is,  one propagating  boson $\x(\T)$, two fermions $\l(\T)$ and $\hat \l(\T)$ and one auxiliary field 
$A(\T)$.  The construction  we will detail    can be generalised to all  other multiplets. The proof can be generalised   to  the  Wess and Zumino  model  with no difficulty but   notational complications.  We   first consider  that $t$~is the real time  and we will shortly shift to the Euclidean time, $t\to it $.   
Call $X(t,\th,\thh)$ the superfield
\bea\nn
X(t,\th,\thh)= \x (t)+     \th   \l  (t)   +   \thh \ll (t)  +  \th\thh   A  (t)
\eea
where $\th$ and $\thh$ are Grassman coordinates, with 
$\int \d\thh    =   \int \d\th     =0$ and 
$\int \d\thh   \thh=   \int \d\th   \th = 1$. Define also 
\bea 
D_\th\equiv  \pa_\th + \ee   \th \pa_t \quad  \quad \quad
D_\thh\equiv    \pa_\thh + \eeh   \thh \pa_t   \cr 
\{D_\th,D_\th\}=\{D_\thh,D_\thh \} =0\quad\quad \quad \demi \{D_\th,D_\thh \}=   \ee\pa_t \eea
The action  of  supersymmetry transformations  on  the superfield $X$ is given by  $Q=  \pa_\th - \ee   \th \pa_t$ and 
$\hat Q=  \pa_\thh -\ee   \thh \pa_t$.
By expansion  in  components,  one finds the following  action of  both  supersymmetries.  \bea\nn
\dd   \x &=&  \l   \quad    \dd \l    = \ee \pa_t  \x    \quad   \dd \ll    =   -A   \quad   \dd A= - \ee \pa_t  \ll       \\
\ddh   \x&=&  \ll    \quad    \ddh \ll    = \eeh \pa_t  \x    \quad   \ddh \l    =   A   \quad   \ddh A=   \eeh \pa_t  \l  
\eea
They     satisfy  $\demi \{    \dd_i,  \dd_j \}=\delta_{ij}\pa_t$ and  $\dd_i\d +\d\dd_i=0$.
The  massless supersymmetric free    action is   
\def\d{{\rm d}}
\bea 
\Sigma_0 \equiv\int \d   t \d  \th    \d \thh  \d \t
(\demi XD_\thh D_\th X)
=\int  \d t
(\demi  A^2-\demi \ee\eeh   \x  \pa_t ^2\x   -\demi \eeh\l \pa_t \l-\demi \ee\ll \pa_t \ll  )
\eea
A general supersymmetric interaction, including a mass term, is  determined by the  prepotential $W(X)$ 
\bea 
\Sigma_W  \equiv\  \int \d   t \d  \th    \d \thh  \d \t
(W(X))
=\int  \d t
(AW_\x(\x) -\l W_{\x\x}(\x)\ll  )
\eea
($W_q,\ W_{qq}$ mean 
$\pa_q W, \ \pa_q^2 W $).
So, the     generic   standard interacting    supersymmetric action    that we consider     is 
\bea\nn
I_W\equiv 
\Sigma _0+ \Sigma_W 
= \int\d t
(\demi  ( A +W_\x)^2   -\demi   W_\x^2
+\demi    { \ee\eeh} (   \pa _t \x)^2   
-\demi  \eeh\l \pa_t \l-\demi  \ee\ll \pa_t \ll   
- \l W_{\x\x}\ll   )
\eea
With  $t\to it$,
one gets the following equivalent (modulo boundary terms)  expressions of the Euclidian action \footnote {As    all  supersymmetric theories,  this model  has a  twisted formulation, where one  can  rewrite the supersymmetry algebra as a nilpotent one. In this case,  one   redefines
$
\psi  =
\frac{\l-  i\ll}{\sqrt{2} }$ and $
=
\frac{\l+ i\ll}{\sqrt{2} i}
$
By doing this change of variables,  the Euclidean action becomes
\def\IT{I_   {  {   \rm{ twisted}}}}   $
{\IT}
= \int\d t
(\demi   A^2      +W_\x A    
-\demi  (\pa_t \x) ^2 
+\bar \psi  \pa_t \Psi 
-\bar \psi W_{\x\x}  \Psi  )
\sim   
\int\d t
(  
-\demi   (\pa_t \x)^2  
-\demi W_\x ^2    
-\bar \psi  \pa_t \Psi 
-\bar \Psi W_{\x\x}  \psi  )
$. One   can reintroduce another auxiliary field $b$, and  get the following twisted formulation for the $N=2$ supersymmetric action, 
${\IT}
=
\int\d t
(  \demi b^2 +b( \pa_t \x+
W_\x)    
-\bar \psi ( \pa_t \psi 
+ W_{\x\x}  \Psi  )  )
$, and the twisted nilpotent supersymmetry is  
$\delta_{\rm twisted} \x=\psi, 
\delta_{\rm twisted}  \psi=0 ,   \delta_{\rm twisted} \bar \psi =b, 
\delta_{\rm twisted} b=0
$, with     ${\IT}
=   \delta_{\rm twisted}
\int\d t
\bar \psi  (   \demi b  + \pa_t \x+
W_\x)
$. In this  expression, the $N=2$~supersymmetry appears twisted, with some interested  properties. As for its generalisation in the stochastic formulation, one can proceed as in the untwisted  case,  and, once one has computed   the stochastic TQFT supersymmetry,  one gets an interesting 4-simplex, because  both   supersymmetries of the    stochastically   quantised action  can be described  as nilpotent ones.}
\bea 
I_W^{t\to it}
&= & \int\d t
(\demi   A  ^2   +   A  W_\x 
-\demi    { \ee\eeh} (\pa_t\x )^2 
- \frac{i}{2}   \eeh\l \pa_t \l- \frac{i}{2}\ee\ll \pa_t \ll   
- \l    W_{\x\x}\ll   )
\cr
&= & \int\d t
(\demi  ( A +W_\x)^2   -\demi   W_\x^2
-\demi    { \ee\eeh} (\pa_t\x )^2  
- \frac{i}{2}   \eeh\l \pa_t \l- \frac{i}{2}\ee\ll \pa_t \ll   
- \l    W_{\x\x}\ll   )
\cr
&= & \int\d t
(\demi  ( A +W_\x)^2   -\demi   (     \pa_t  \x  +W_\x)^2
- \frac{i}{2}   \eeh\l \pa_t \l- \frac{i}{2}\ee\ll \pa_t \ll   
- \l    W_{\x\x}\ll   )
\eea\def\IM{I_{\small w}}
From now on we will  denote 
$\IM    \equiv  -I_W^{t\to it} $, keeping in mind that we compute Euclidean correlation functions in the $\tau=\infty$ limit.  
In order to achieve the  stochastic quantisation,  one uses the Euclidean action.  It  is unclear if  the Wick rotation  on $t$ can be done at finite values of the stochastic time.
The possibility of a Wick rotation can be however checked in the  limit $\tau=\infty$.

For  polynomial interactions, it is convenient to separate the mass  term $\demi m^2 \x^2$ from the rest of the interactions  in $W(q)$, which are of degree  higher than 2, with  $
W  \equiv  \demi m \x^2  +V(\x) $ and  $m=W_{\x\x}(0)
$. This decomposition will be  used shortly,  to   check perturbatively   the    possible  stochastic equilibrium of the supersymmetric  model.

We must define the  supersymmetry covariance  of all   fields of the stochastic process. Consider the 
  noise  fields $\eta_a$.  
Given  a   component  field $a$  of  the multiplet $X$, its noise $\eta_a$    is   a  random fluctuations of $a$,  modulo terms proportional to some equations of motion.  Thus, 
if  stochastic  quantisation    is compatible with supersymmetry,   the noises of 
$\x,A, \l,\ll$      build      a superfield    
\bea
{\bf \eta}=   \eta_\x    +\th   \eta _l    + \thh  \eta_\hl   +\th\thh   \eta_A 
\eea
and transform accordingly.
As we will demonstrate,   non-trivial   kernels $K_{ab} $ are necessary in order to obtain      supersymmetry covariant   Langevin equations as well as   
a proper convergence of the Langevin process. The Langevin equation that respects supersymmetry must be written as
\bea
\dot \x  &=& - \sum_{a=\x,\l_i,b}  K_{\x,a} \frac {\delta {\IM}}{\delta a} +\eta_\x   \cr
\dot A  &=&   - \sum_{a=\x,\l_i,b}    K_{A,a} \frac {\delta {\IM}}{\delta a} +\eta_A  \cr
\dot  \l&=&   - \sum_{a=\x,\l_i,b}  K_{\l,a} \frac {\delta {\IM}}{\delta a} +\eta_\l   \cr
\dot \ll  &=&  - \sum_{a=\x,\l_i,b}  K_{\ll ,a} \frac {\delta {\IM}}{\delta a} +\eta_\ll     \
\eea
We    will  use   superfield arguments
to compute the appropriate kernels $K_{ab} $. 

\section {Determination of the kernels to  enforce  supersymmetry at  all values of stochastic time}

We will use  the following remark :     $X$ being  a superfield,     the supersymmetric  invariance of the action ${\IM} (X) $ implies  that  its equations of motion build the following "dual" superfield 
\bea
X^*_W=  \frac{ \delta {\IM}} {\delta A }
  -\th  \frac{ \delta {\IM}}{ \delta \ll }     +  \thh \frac{ \delta {\IM}}{ \delta \l }      +  \th\thh  \frac{ \delta {\IM}} {\delta \x },      
\eea
independently of the  details of   the supersymmetric action and thus of   $W$.    The proof is simple:  if one varies $X\to X +\delta X$,  the variation of   ${\IM}$ 
is   the supersymmetric term $\int dx
( \frac{ \delta {\IM}} {\delta \x }  \delta \x+
    \frac{ \delta {\IM}}{ \delta \l}   \delta\l+
  \frac{ \delta {\IM}}{ \delta \ll }   \delta\ll+
\frac{ \delta {\IM}} {\delta A }  \delta A)
$.  This   variation is nothing but    the  integral over $\d\th\d\thh$ of the  superfield $\delta X$ times $  X^*_W$, which picks out the coefficients  term of  $\th\thh$
from the product $  X^*_W \delta X $.
Thus $  X^*_W$  is a superfield, and one has:
\bea
X^*_W=  -A-W_q
+\th(-i\pa_t \ll + W_{qq}\l)
-\thh(-i\pa_t \l-W_{qq}\ll)
 - \th\thh (   \pa_t ^2q +A  W_{qq}-\l W_{qq}\ll)     
\eea
The stochastic time derivative of $X$  is  also a superfield, 
\bea
\dot  X(t,\th,\thh,\t)& =&    \dot \x +     \th  \dot \l   + \thh \dot\ll    +  \th\thh  \dot A 
\eea
 We will need more independent superfields to define  the  stochastic  BRST supersymmetry framework.

Let   $a, b,...$ denote the components $x,\l,\ll,A$  of the superfield $X$.
The goal is to write a supersymmetric action   of the form 
$\s    {{\bar \Psi} _a}   [      K_{a b}b _b+
\frac {\d  X a   } {\d \tau } +K_{ab}  \frac {\delta  {  S} } {\delta X_ b}   ]$,  
where $\s$ is the stochastic BRST symmetry operation we defined for    Eq.~(\ref{stoctruc}).  In the case we discuss, one has  $\s X_a =\Psi_a, \ \s \Psi_a =0,\ 
\s\bar  \Psi_a  =b_a, \ \s  b_a   =0$. 
We will express the $\s$-exact action depending on   $X$ and its  BRST topological partners  as  the integral of a super-Lagrangian. 
\def\mb{{\blue \xi} m}
\def\mb{ M}

Then, we will integrate this super-Lagrangian  over superspace $(\th,\thh)$ and get the  supersymmetric $\s$-exact action in components. Using   Eq.~(\ref{stoctruc}), we will identify  in a reversed way the Langevin equation for each component in $X$. Indeed,
 they are encoded in 
Eq.~(\ref{stoctruc})   as the coefficient  of the terms linear in the $b$'s. Such Langevin equations will be covariant under supersymmetry, by construction.

To impose  the supersymmetry transformation of the $b_a$'s, we simply  declare  that
\bea
B\equiv  b_A+ \th b_\ll    +\thh b_\l  +\th\thh b_\x
\eea
is a  superfield. This   determines the supersymmetry transformations of all components $b_a$.

Thus, the program is to first  write a super-action depending on $B$,  which has   the form of the $b$-dependant terms  in~Eq.(\ref{stoctruc}) after integration over superspace.  

To get  the stochastic ghost depending part of ~Eq.(\ref{stoctruc}), and  enforce the compatibility of the supersymmetry of our model with the stochastic quantisation, we also  introduce  the   stochastic antighost superfield  $\bar \Psi _X$, with
$\s \bar \Psi_X=B$, $\s B=0 $, 
\bea
\bar \Psi_X \equiv  \bar \Psi_A+ \th \bar \Psi_\ll    +\thh \bar \Psi_\l  +\th\thh \bar \Psi_\x
\eea
and the  stochastic  ghost superfield $\Psi_X$ of  $X$,with  $\s X= \Psi_X$,   $\s  \Psi_X=0 $ and  the decomposition 
\bea
\Psi _X\equiv  \Psi_\x + \th  \Psi_\l    +\thh   \Psi_\ll +\th\thh \bar \Psi_A
\eea
For the sake of the definition of the kernels, we  now introduce      a mass parameter $\mb$. This  parameter    may have to  be  fine-tuned, to ensure the eventual convergence of the stochastic process, using the kernels that we shall shortly determine. The values of the correlators at  $\tau=\infty$, if the limit exists,  
will not  depend on the choice  of $\mb$ according to the general proof of the $K$ independence of the limit.

One then   defines 
 the   following  supersymmetric quadratic functional of B :  
\bea
I_\mb  &=& 
\int \d   t \d  \th    \d \thh  \d \t
(\demi B D_\thh D_\th B- \demi    \mb B^2)\cr
&=& 
\int\d t
(
\demi    b_\x ^2 -\mb b_\x b_A 
-  \frac {1}{2} (  \pa_t b_A)^2 
-  \frac {i}{2}       b_\l    \pa_t     b_\l
- \frac {i}{2}    b_\ll   \pa_t    b_\ll   
+   \mb b_\ll  b_\l
) 
\eea
Whatever  the value of   $\mb$ is in this supersymmetric auxiliary free action,       the   equations of motion of $B$    determine  the "dual"  superfield : 
\bea
B_{I_{\mb}}^*&\equiv & b_\x - \mb b_A     
+\th  (i \pa_t b_\l +\mb b_\ll )  
+\thh(-i \pa_t  b_\ll   +  \mb b _\l  )
 +  \th\thh  ( \pa_t^2 b_A    -\mb b_\x)   
\eea
%
Since     $\dot X ,    
X^*_W, B, B_{I_{\mb}}^*$  are  all  superfields, the  following action is supersymmetric :
\bea
I_{stoc, B}^{\mb}&=&
\int \d t\d\th \d\thh \d \t  \Big [
\demi BB_{I_{\mb}}^* + B  \dot X  +B_{I_{\mb}}^*   X^*_W
\Big ] 
\eea
It can be expressed in components as 
\bea
I_{stoc, B}^{\mb}
&=&
\int \d t   \d\t  \Big [
\demi b_\x ^2
+
\demi b_A\pa^2_t b_A 
-\mb b_\x b_A 
-\frac{i}{2}     ( b_\l   \pa_t b_\l 
+b_\ll  \pa_t b_\ll) -M  b_\ll b_L
\cr
& &
+b_\x
( \dot \x + \frac{ \delta {\IM}} {\delta \x }    - \mb   \frac{ \delta {\IM}} {\delta A }) 
+  b_A
( \dot  A +\pa_t^2  \frac{ \delta {\IM}} {\delta A }    -\mb \frac{ \delta {\IM}} {\delta \x }   )
\cr
& &
+b_\l(\dot  \l  +  i\pa_t   \frac{ \delta {\IM}}{ \delta \l }    -\mb \frac{ \delta {\IM}}{ \delta \ll }  )  
-b_\ll(\dot  \ll   +i \pa_t  \frac{ \delta {\IM}}{ \delta \ll }    + \mb \frac{ \delta {\IM}}{ \delta \l  })
)\Big]
\cr
&=&
\int \d t   \d\t   \Big [
\demi {b'_\x } ^2
+\demi b_A(\pa^2_t -\mb ^2) b_A 
-\frac{i}{2}     ( b_\l   \pa_t b_\l 
+b_\ll  \pa_t b_\ll) -\mb b_\ll b_\l
\cr
& &
+ b'_\x 
\Big ( \dot \x  + \frac{ \delta {\IM}} {\delta  \x}    - \mb   \frac{ \delta {\IM}} {\delta A } \Big)
\cr & &
+  b_A
\Big  ( \dot  A +(\pa_t^2 -\mb ^2)
\frac{ \delta {\IM}} {\delta A } 
+\mb( \dot \x +  \frac{ \delta {\IM}} {\delta  \x}  
- \mb\frac{ \delta {\IM}} {\delta  A} 
)
\Big)
\cr
& &
+b_\l  \Big   (\dot  \l  + i\pa_t   \frac{ \delta {\IM}}{ \delta \l }    -\mb \frac{ \delta {\IM}}{ \delta \ll }  \Big)  
-b_\ll\Big(\dot  \ll  +i \pa_t  \frac{ \delta {\IM}}{ \delta \ll }    +\mb \frac{ \delta {\IM}}{ \delta \l  } 
\Big)
\Big]
\eea
where $b_x'=b_\x -\mb b_A$.  
This supersymmetric  action  has a quadratic dependance 
in the fields   $b_a =(b_ \x,b_\l,b_\ll, b_A)$, which are  the auxiliary fields of 
the topological BRST symmetry of stochastic quantisation, as in Eq.~(\ref{stoctruc}). The     coefficients  of the linear terms   in the $b$'s  provide the supersymmetry covariant       Langevin equations
\bea
\dot  \x   &=&  - \frac{ \delta {\IM}} {\delta \x }    + \mb   \frac{ \delta {\IM}} {\delta A } +\eta_\x
\cr  
\dot \l   &=& 
-  i \pa_t \frac{ \delta {\IM}}{ \delta \l }      + \mb  \frac{ \delta {\IM}}{ \delta \ll }     +\eta_\l
\cr  
\dot \ll   &=& 
-i \pa_t \frac{ \delta {\IM}}{ \delta \ll }      - \mb  \frac{ \delta {\IM}}{ \delta \l}       +\eta_\ll
\cr
\dot A    &=&-
(\pa_t^2-2 \mb^2)   \frac{ \delta {\IM}} {\delta A } 
-\mb( \dot \x  + \frac{ \delta {\IM}} {\delta \x }    )  +\eta_A
 = 
-( \pa_t^2 -\mb^2)  \frac{ \delta {\IM}} {\delta A }  +\eta_A
-\mb \eta_\x
\eea
The quadratic terms  in $b$ define the  Gaussian noise distribution   by
\bea
<<f(\eta)>>=  \int [\d\eta_q]
[\d\eta_\l]
[\d\eta_\ll]
[\d\eta_A]\ 
f(\eta)   \quad\quad\quad\quad\quad\quad \quad \cr
\exp-\demi
\int \d t \d\tau \Big ( \eta_q^2
+\eta_A\frac{1}{\pa_t ^2  -M^2}\eta_a
+ \demi \eta_\l    \frac {i\pa_t}{\pa_t ^2-  M^2} \eta_\l    
+ \demi \eta_\ll    \frac {i\pa_t}{\pa_t ^2  -M^2} \eta_\ll  +\eta_\l    \frac {M}{\pa_t ^2  -M^2} \eta_\ll  \Big)
\eea
The general theorem applies and the limit of the field correlators  in the limit $\tau\to \infty$ is independent on the chosen value of M.

To see perturbatively  the  $\tau\to\infty$ convergence of the solutions of Langevin equations,     one expands       
$W_{qq} (q) =  {m}{ } +V_{qq}(q)$ to  verify that  the drift forces are   negative  for all fields for  
$W= \frac {m}{2}\x^2$ \footnote{The stochastic equation for $A$ is consistent    if one eliminates $A$ by its equation of motion, $A=-m\x$, so there is no evolution for $A$ provided $\eta_A=M\eta_\x$. But the the various closure relations hold modulo equations of motion.}
\bea\label{lasu}
\dot  \x   &=&  (\pa_t^2-m\mb) q +(m-M) A  +\eta_\x
\cr  
\dot \l   &=& 
(\pa_t^2-mM) \l -i (m-M)  \pa_t \ll   +\eta_\l
\cr  
\dot \ll   &=& 
(\pa_t^2-mM) \ll +i (m-M)  \pa_t \l   +\eta_\ll
\cr
\dot A    &=&
(\pa_t^2  -\mb^2) (A+mq)     +\eta_A
- \mb \eta_\x  \eea
Both  operators $\pa_t^2 -m\mb$ and   $\pa_t^2 - \mb^2$ have  negative   eigenvalues in the Fourier transform space. (This requires the   Euclidian formulation for the  stochastic quantisation).  One must  verify    the  the  negativity of the 
eigenvalues of the  differential operators that define the  $\tau$ evolution of      $A,\x$ and   $\l,\ll $. They are 
\bea
\begin{pmatrix}
\pa_t^2 -m\mb  &    m-\mb      \\
m(\pa_   t^2 -\mb^2 )     &  \pa_t^2 - \mb ^2
\end{pmatrix}\quad {\rm  and} \quad
\begin{pmatrix}
\pa_t^2 -m\mb  &   -i( m-\mb) \pa_t      \\
i( m-\mb) \pa_t  &  \pa_t^2 -m\mb
\end{pmatrix},\nn
\eea 
The eigen values  are non-degenerate  for  $\mb \neq m$ and degenerate for  the  simplest and easiest  choice  $\mb=  m$. The determinant of the first  matrix is   $(\pa_   t^2 -\mb^2 )(\pa_t^2 -m ^2 )>0$ and for the second matrix it is  $(\pa_t^2 -m\mb)^2 -(m-\mb)^2 \pa_t^2>0
$.  Moreover the traces of both  matrices are   negative. Thus, whatever the value of   $M$ is, 
  we have the non positivity requirement and a normalisable vacuum for the  supersymmetric Fokker--Planck process, and 
 the stochastic process is converging at $\tau \to \infty$ for the fields $\x,A, \l,\ll$.
The values of correlators at $\tau \to \infty$ is independent on the choice of $\mb$, although the details of the evolution are $\mb$ dependent. The solutions have a different    dependence on $\tau$ for
$\mb \neq m$ and   $\mb=  m$, but the limit $\mb\to m$ is continuous.

We  have thus obtained a set of Langevin equations  with a well defined convergence at $\t=\infty$ that are by construction covariant under the $N=2$ supersymmetry. 
For $M=m$,  the stochastic process is simplest because no mixing occurs between the drift forces of all fields at the  free level, and  the Fokker--Planck Hamiltonian only involves   the   eigenvalues of  the single operator  $-\pa_t^2+\mb^2 $ for all  fields.  

We can now  write the complete $\s$-exact action associated to these supersymmetric  Langevin equations, that is, a  stochastic  BRST-TQFT  supersymmetric  action in the 2d-space $\{t,\tau\}$,  which  is   the integral of a superfield,  and thus  a  $N=2$   supersymmetric  2d  action.  

We use the   already anticipated stochastic topological superghosts  $\Psi_X$ and  $\bar \Psi _X $, 
upon which  the graded differential operator 
$\s$ acts as  
$\s X =\Psi_X,
\  
\s  \Psi_X=0,\  
\s \bar \Psi _X =B,\
\s  B=0
$. Defining,
\bea {\bar \Psi}_{I_{\mb}}^*\equiv   {\bar \Psi}_\x - \mb {\bar \Psi}_A     
+\th  (i \pa_t {\bar \Psi}_\l +\mb {\bar \Psi}_\ll )  
+\thh(-i \pa_t  {\bar \Psi}_\ll   +  \mb {\bar \Psi} _\l  )
 +  \th\thh  ( \pa_t^2 {\bar \Psi}_A    -\mb {\bar \Psi}_\x) 
 \eea
 one has 
  $\s  {\bar \Psi}_{I_{\mb}}^*   =   B_{I_{\mb}}^*$ and $ \s  B_{I_{\mb}}^*=0$. 
The  complete  N=2 supersymmetric action that is    invariant under the stochastic BRST supersymmetry and expresses the Langevin equations~(\ref{lasu})  is  thus
\bea
I_{stoc}&=&
\int \d t\d\th \d\thh  \d\t   \s    \Big [\bar \Psi_{I_{\mb}}^* (\demi  B +  X^*_W
+\bar \Psi _X\dot X)
\Big ]
\eea
It can  be  simply expanded in components. Then, all the details of the stochastic evolution of the correlators of all components  of the superfield    can be studied while maintaining the supersymmetry.  
\section{Conclusion}
Using a superfield construction,  we explained  the way the  stochastic time evolution of  the stochastically quantised  $N=2$ supersymmetric  quantum mechanics   correlation functions  is compatible with the   $N=2$ supersymmetry. The stochastic   TQFT  supersymmetry       and the  space supersymmetry  satisfy
\bea
[\s,\frac{\d}{\d \tau}]=0 \quad
\{ \s,    Q \}=0,
\quad
\{\s,\hat Q \}=0
\eea
for every finite value of the stochastic time, which is the required property.   We  introduced kernels  that  ensure the convergence of the stochastic process toward  the standard path integral of $N=2$ supersymmetric quantum mechanics.  These kernels are covariant under supersymmetry and depend on a massive parameter that can be chosen at will.  Its value modifies the course of the stochastic quantisation, but the limit of the correlation functions when  the stochastic time runs to $\tau=\infty$ is independent on its choice.

The same demonstration can be mutatis mutandi  repeated for the Wess and Zumino model. 
It is also possible to repeat  this construction  to more sophisticated supersymmetric theories, including those with a gauge invariance, provided one uses the method of equivariant cohomology.

 \bigskip\noindent {\bf Acknowledgment.}
 I wish  to thank 
 the NCTS in Hsinchu  as well as the Max Planck Institute  in Golm  for their  generous and warm hospitality.

\end{document}